\documentclass[sigconf]{acmart}

\usepackage[utf8]{inputenc}
\usepackage{amsmath}
\usepackage{booktabs}
\usepackage{enumitem}
\usepackage{graphicx}
\usepackage{subcaption}
\usepackage{listings}
\usepackage{xcolor}
\usepackage{hyperref}
\usepackage{float}
\usepackage{tikz}
\usepackage{pgfplots}
\pgfplotsset{compat=1.18}
\usetikzlibrary{shapes,arrows,positioning,fit,backgrounds}

\lstdefinelanguage{Rust}{
  keywords={fn, let, mut, pub, use, mod, struct, enum, impl, trait,
             for, while, loop, if, else, match, return, unsafe, extern,
             crate, super, self, Self, type, where, const, static, ref,
             move, box, in, continue, break, as, true, false, dyn, async,
             await, yield, str, bool, u8, u16, u32, u64, u128, usize,
             i8, i16, i32, i64, i128, isize, f32, f64, char},
  keywordstyle=\color{blue},
  sensitive=true,
  comment=[l]{//},
  morecomment=[s]{/*}{*/},
  commentstyle=\color{gray},
  stringstyle=\color{red!70!black},
  string=[b]",
  morestring=[b]',
}
\definecolor{bugOrange}{RGB}{230,110,0}
\definecolor{bugPurple}{RGB}{140,0,180}
\definecolor{bugTeal}{RGB}{0,140,140}
\definecolor{bugRed}{RGB}{200,0,0}
\definecolor{safeGreen}{RGB}{0,140,60}

\newcommand{\hc}[2]{{\color{#1}\textbf{#2}}}

\newcommand{\hlOrange}[1]{\hc{bugOrange}{#1}}
\newcommand{\hlPurple}[1]{\hc{bugPurple}{#1}}
\newcommand{\hlTeal}[1]{\hc{bugTeal}{#1}}
\newcommand{\hlRed}[1]{\hc{bugRed}{#1}}
\newcommand{\hlGreen}[1]{\hc{safeGreen}{#1}}

\copyrightyear{2026}
\acmYear{2026}
\setcopyright{acmlicensed}
\acmConference[SSE '26]{The 6th International Workshop on Software Security Engineering}{June 9--12, 2026}{Glasgow, Scotland, United Kingdom}
\acmBooktitle{Proceedings of the 6th International Workshop on Software Security Engineering (SSE '26), co-located with EASE 2026, June 9--12, 2026, Glasgow, Scotland, United Kingdom}
\acmDOI{10.1145/XXXXXXX.XXXXXXX}
\acmISBN{978-1-4503-XXXX-X/26/06}

\acmSubmissionID{12}

\begin{document}

\title{Symbolic Execution Meets Multi-LLM Orchestration: Detecting Memory Vulnerabilities in Incomplete Rust CVE Snippets}

\author{Zeyad Abdelrazek}
\authornote{Corresponding author.}
\email{zgad01@jaguar.tamu.edu}
\affiliation{
  \institution{Texas A\&M University--San Antonio}
  \streetaddress{One University Way}
  \city{San Antonio}
  \state{Texas}
  \postcode{78224}
  \country{USA}
}

\author{Young Lee}
\email{ylee@tamusa.edu}
\affiliation{
  \institution{Texas A\&M University--San Antonio}
  \streetaddress{One University Way}
  \city{San Antonio}
  \state{Texas}
  \postcode{78224}
  \country{USA}
}

\begin{abstract}
This paper presents a system combining symbolic execution (KLEE) with a 4-agent multi-LLM architecture for detecting memory vulnerabilities in Rust \texttt{unsafe} code. A central challenge we address is the \emph{incomplete-code problem}: CVE database entries provide only isolated code snippets that lack struct definitions, imports, and Cargo manifests, causing all existing formal verification tools to fail at compilation with zero output. Our system resolves this through four specialized agents---an Oracle/Validator for strategic planning, a Safety Checker for vulnerability analysis, a Code Specialist for FFI wrapper generation, and a Fast Filter for execution optimization---that collaboratively synthesize KLEE-compatible harnesses from otherwise uncompilable fragments. KLEE's output is then ingested by \texttt{graph\_klee.py}, which constructs a Graph Database linking CVE files, CWE categories, error types, and symbolic execution paths as typed nodes and labelled edges, enabling structured cross-CVE vulnerability queries. We evaluated our system on 31 real-world Rust CVEs spanning 11 CWE categories, achieving 90.3\% wrapper compilation success where all state-of-the-art formal verification tools achieve 0\%. Our system detected 1,206 critical errors across 26 files (83.9\% detection rate), compared to 14 warnings across 11 files for Clippy (35.5\%) and generic labels for Miri. The 4-agent architecture reduced wrapper compilation failures from 42\% (single-agent baseline) to 9.7\% and increased detected errors from 487 to 1,206, confirming that role specialization and structured context passing produce measurably better results than a single general-purpose model. Our replication package is publicly available at \url{https://github.com/Zeyad-Ab/Symbolic-Execution-with-Multi-LLM-Architecture-for-Rust-Security}.
\end{abstract}




\maketitle

\section{Introduction}
\label{sec:Introduction}

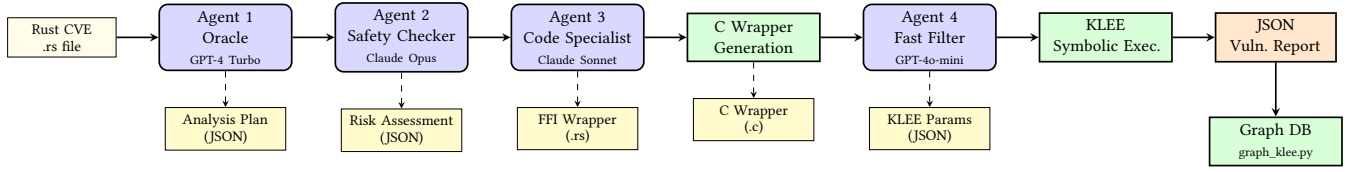
\begin{figure*}[!t]
\centering
\resizebox{\textwidth}{!}{
\begin{tikzpicture}[
  node distance=0.4cm and 0.7cm,
  agent/.style={rectangle, draw, fill=blue!15, text width=2cm, align=center,
                minimum height=0.9cm, rounded corners, thick, font=\small},
  artifact/.style={rectangle, draw, fill=yellow!25, text width=1.8cm, align=center,
                   minimum height=0.55cm, font=\footnotesize},
  process/.style={rectangle, draw, fill=green!15, text width=2cm, align=center,
                  minimum height=0.8cm, thick, font=\small},
  output/.style={rectangle, draw, fill=orange!20, text width=1.8cm, align=center,
                 minimum height=0.7cm, thick, font=\small},
  data/.style={rectangle, draw, fill=yellow!10, text width=1.6cm, align=center,
               minimum height=0.55cm, font=\footnotesize},
  arrow/.style={->, thick, >=stealth},
  darrow/.style={->, dashed, >=stealth}
]

\node[data]    (input)  {Rust CVE\\.rs file};
\node[agent,   right=of input]  (a1)     {Agent 1\\Oracle\\{\scriptsize GPT-4 Turbo}};
\node[agent,   right=of a1]     (a2)     {Agent 2\\Safety Checker\\{\scriptsize Claude Opus}};
\node[agent,   right=of a2]     (a3)     {Agent 3\\Code Specialist\\{\scriptsize Claude Sonnet}};
\node[process, right=of a3]     (cwrap)  {C Wrapper\\Generation};
\node[agent,   right=of cwrap]  (a4)     {Agent 4\\Fast Filter\\{\scriptsize GPT-4o-mini}};
\node[process, right=of a4]     (klee)   {KLEE\\Symbolic Exec.};
\node[output,  right=of klee]   (report) {JSON\\Vuln. Report};
\node[process, below=0.9cm of report] (graphdb) {Graph DB\\{\scriptsize graph\_klee.py}};

\node[artifact, below=0.6cm of a1]    (plan)   {Analysis Plan\\(JSON)};
\node[artifact, below=0.6cm of a2]    (risk)   {Risk Assessment\\(JSON)};
\node[artifact, below=0.6cm of a3]    (ffi)    {FFI Wrapper\\(.rs)};
\node[artifact, below=0.6cm of cwrap] (cwfile) {C Wrapper\\(.c)};
\node[artifact, below=0.6cm of a4]    (params) {KLEE Params\\(JSON)};

\draw[arrow] (input)  -- (a1);
\draw[arrow] (a1)     -- (a2);
\draw[arrow] (a2)     -- (a3);
\draw[arrow] (a3)     -- (cwrap);
\draw[arrow] (cwrap)  -- (a4);
\draw[arrow] (a4)     -- (klee);
\draw[arrow] (klee)   -- (report);
\draw[arrow] (report) -- (graphdb);

\draw[darrow] (a1)    -- (plan);
\draw[darrow] (a2)    -- (risk);
\draw[darrow] (a3)    -- (ffi);
\draw[darrow] (cwrap) -- (cwfile);
\draw[darrow] (a4)    -- (params);

\end{tikzpicture}
 }
\caption{\textbf{4-Agent Pipeline Overview.} The pipeline proceeds left to right through six stages. (1)~A Rust CVE snippet (incomplete source file) is fed to \textbf{Agent~1 (Oracle/Validator, GPT-4 Turbo)}, which identifies likely vulnerability types and produces a structured JSON analysis plan---constraining all downstream agents to the relevant vulnerability classes. (2)~\textbf{Agent~2 (Safety Checker, Claude Opus)} receives the snippet and the plan, performs deep security analysis, assigns a risk score (0--10), and annotates critical lines. (3)~\textbf{Agent~3 (Code Specialist, Claude Sonnet)} uses all prior outputs to generate a complete, compilable Rust FFI wrapper that replicates the vulnerability class in KLEE-compatible types; an automated step then synthesizes the corresponding C harness. (4)~\textbf{Agent~4 (Fast Filter, GPT-4o-mini)} selects optimal KLEE parameters (search strategy, time/memory limits, fork depth) from the risk score. (5)~KLEE performs symbolic execution on the compiled wrapper and emits a structured JSON vulnerability report. (6)~\textbf{graph\_klee.py} ingests the JSON report and constructs a \textbf{Graph Database} in which nodes represent CVE files, CWE categories, error types, and symbolic paths, and edges encode the relationships between them---enabling cross-CVE vulnerability queries, pattern clustering, and structured export for downstream analysis. Dashed arrows show intermediate artifacts persisted to disk for reproducibility.}
\label{fig:methodology}
\end{figure*}

Rust is one of the fastest-growing systems programming languages, valued for its ownership-based memory safety guarantees~\cite{rust_ownership}. Yet these guarantees do not extend to \texttt{unsafe} code blocks, which are necessary for low-level operations and Foreign Function Interfaces (FFI) but introduce the same class of memory errors---buffer overflow, use-after-free, double-free---that plague C and C++. The practical consequence is a growing catalogue of Rust CVEs whose \texttt{unsafe} code contains real, exploitable memory faults.

A critical and largely overlooked challenge compounds this problem: CVE database entries provide only isolated code \emph{snippets} stripped of struct definitions, trait implementations, imports, and Cargo manifests. This \emph{incomplete-code problem} creates a hard barrier for every existing tool. KLEE~\cite{klee_original} operates on LLVM bitcode from C/C++ and requires a compiled FFI harness that Rust snippets cannot provide. Clippy~\cite{rust_book} matches syntactic patterns and misses path-dependent bugs entirely. Formal tools (Kani~\cite{kani}, Prusti~\cite{prusti_modular}, Creusot~\cite{creusot}, Haybale~\cite{haybale}) require a complete Cargo project with resolved type information---applied to CVE snippets, all fail at compilation with zero output. The result is that no automated tool today can perform deep, execution-level vulnerability analysis on the incomplete Rust code CVE databases actually contain.

This paper closes that gap. We introduce a 4-agent multi-LLM pipeline that transforms an incomplete Rust CVE snippet into a KLEE-compatible FFI harness, runs symbolic execution, and constructs a Graph Database from the results. \textbf{To the best of our knowledge, this is the first system to enable symbolic execution on incomplete, context-free Rust CVE snippets without manual intervention}, achieving 90.3\% wrapper compilation success where every existing formal tool achieves 0\%.

This paper addresses the following research questions:

\textbf{RQ1:} \textit{Can symbolic execution, enabled by LLM-generated FFI wrappers, detect memory vulnerabilities in incomplete Rust CVE snippets?} Specifically, we ask whether KLEE can flag at least one critical error in each CVE file across 11 CWE categories.

\textbf{RQ2:} \textit{To what extent can LLM-generated FFI wrappers bridge the incomplete-code gap and enable symbolic execution on isolated Rust CVE snippets?} We measure compilation success rate compared to 0\% for formal verification tools.

\textbf{RQ3:} \textit{Does a specialized 4-agent LLM architecture outperform a single-agent approach in vulnerability detection rate, wrapper compilation success, and error count?}

Our system employs four distinct agents: an Oracle Validator (GPT-4-Turbo) for strategic planning, a Safety Checker (Claude Opus 4.5) for deep security analysis, a Code Specialist (Claude Sonnet 4.5) for generating FFI wrappers, and a Fast Filter (GPT-4o-mini) for optimizing KLEE parameters. A post-processing component, \texttt{graph\_klee.py}, then constructs a Graph Database from KLEE's output to enable structured cross-CVE vulnerability queries. Our replication dataset, CVE files, generated wrappers, and experiment scripts are available at \url{https://github.com/Zeyad-Ab/Symbolic-Execution-with-Multi-LLM-Architecture-for-Rust-Security}.

\noindent\textbf{Contributions.} This paper makes the following contributions:
\begin{itemize}[leftmargin=*,itemsep=2pt]
  \item \textbf{First incomplete-code symbolic execution pipeline for Rust.} To our knowledge, this is the first automated system capable of running KLEE symbolic execution on isolated Rust CVE snippets that lack struct definitions, imports, and Cargo context---achieving 90.3\% compilation success where all state-of-the-art formal verification tools achieve 0\% and no prior system has reported any success.
  \item \textbf{First multi-agent LLM architecture for Rust CVE analysis.} A coordinated 4-agent pipeline in which role-specialized models collaboratively synthesize KLEE-compatible FFI wrappers from uncompilable Rust fragments---reducing wrapper compilation failures from 42\% (single-agent) to 9.7\%, a combination not previously attempted in the literature.
  \item \textbf{Empirical evaluation on real-world CVEs.} A systematic study on 31 real-world Rust CVEs spanning 11 CWE categories, detecting 1,206 critical errors at an 83.9\% detection rate---26.8$\times$ more issues than Miri and Clippy combined, with zero false negatives relative to Clippy.
  \item \textbf{Graph Database post-processing.} A \texttt{graph\_klee.py} component that automatically ingests any KLEE output directory and constructs a Graph Database in which CVE files, CWE categories, error types, and symbolic execution paths are represented as typed nodes and labelled edges---enabling structured cross-CVE vulnerability queries, CWE-level pattern clustering, and export for downstream security analysis.
  \item \textbf{Memory-safety CVE subset and replication package.} A filtered subset of 31 memory-safety CVEs drawn from the publicly available HALURust dataset~\cite{halurustdataset}, annotated with ground-truth CWE labels, together with all generated FFI wrappers, KLEE outputs, Graph DB exports, and replication scripts released publicly.
\end{itemize}

\section{Motivating Example}
\label{sec:motivation}

\subsection{The Vulnerability: CVE-2020-35904}

Consider CVE-2020-35904, a confirmed memory-corruption vulnerability in the \texttt{concurrent-queue} crate (CWE-131: Incorrect Calculation of Buffer Size). The CVE database provides the snippet shown in Listing~\ref{lst:cve_motivation}: a \texttt{drop} implementation that reconstructs a \texttt{Vec} with \texttt{length=0} while passing the actual capacity, causing the allocator to free an incorrectly sized region. The \texttt{unsafe} block on the final line is the root cause---but to understand \emph{why} it is dangerous, an analysis tool must be able to execute or symbolically explore the function. As we show next, no existing tool can do this without help.

\begin{lstlisting}[language=Rust,
  caption={CVE-2020-35904 as it appears in the vulnerability database.
  The struct fields (\texttt{self.head}, \texttt{self.cap},
  \texttt{self.buffer}) and the \texttt{Ordering} import are absent,
  causing immediate compilation failure in all existing tools.},
  label={lst:cve_motivation}]
fn drop(&mut self) {
    // self.head undefined -- struct not present
    let hix = self.head.load(Ordering::Relaxed)
              & (self.one_lap - 1);
    for i in 0..self.len() {
        let index = if hix + i < self.cap {
            hix + i
        } else { hix + i - self.cap };
        unsafe {
            let slot = &mut *self.buffer.add(index);
            let value = &mut *slot.value.get();
            value.as_mut_ptr().drop_in_place();
        }
    }
    // BUG: length=0 causes incorrect deallocation
    unsafe { Vec::from_raw_parts(self.buffer, 0, self.cap); }
}
\end{lstlisting}

\subsection{Why KLEE Cannot Analyze This Directly}

KLEE operates on LLVM bitcode compiled from C/C++ and has no native Rust support~\cite{klee_original}. Applying KLEE to a Rust function requires two preconditions: the code must compile with \texttt{rustc} to LLVM bitcode, and a C FFI harness must exist calling \texttt{klee\_make\_symbolic()} on all inputs. Neither is met here. The fields \texttt{self.head}, \texttt{self.cap}, and \texttt{self.buffer} belong to a struct entirely absent from the snippet; \texttt{Ordering} is unimported. Running \texttt{rustc} produces immediate \emph{undeclared type} errors---no bitcode, no harness, no KLEE.

\subsection{Why Clippy Cannot Detect the Bug}

Clippy matches syntactic and type-based \emph{patterns}---integer casts, unreachable code, suspicious pointer operations. The bug here is \emph{semantic and path-dependent}: \texttt{Vec::from\_raw\_parts} is called with \texttt{length=0} instead of the actual length, which is only wrong because of its pairing with \texttt{cap}. No lint rule can detect this without reasoning about the relationship between length and capacity at runtime---precisely the kind of reasoning that requires execution-level analysis. Clippy reported \textbf{zero warnings}.

\subsection{Why All Other Tools Fail on Incomplete Code}

Beyond KLEE and Clippy, every Rust analysis tool requires \emph{complete, compilable code}. \textbf{Formal verification tools} (Kani~\cite{kani}, Prusti~\cite{prusti_modular}, Creusot~\cite{creusot}, Haybale~\cite{haybale}) each build on the Rust compiler pipeline and require a complete crate with a \texttt{Cargo.toml}---Creusot further rejects \texttt{unsafe} blocks entirely. We submitted CVE-2020-35904 to all four; all failed immediately with missing-type errors and produced zero output. \textbf{Cargo ecosystem tools} (cargo-audit, cargo-geiger, cargo-deny) operate at the manifest level and cannot parse a bare \texttt{.rs} file. \textbf{Miri} requires a compilable, runnable program and fails at the same compilation step.

The pattern is universal: every existing Rust analysis tool is designed for \emph{complete programs}, not CVE snippets.

\section{Related Work}
\label{sec:related}

\subsection{Formal Verification and Practical Analysis in Rust}
Formal verification tools offer strong correctness guarantees but face significant limits for real-world CVE analysis. Kani~\cite{kani} uses bounded model checking but requires fully compilable code and complete type information. Prusti~\cite{prusti_modular} relies on deductive verification through manual specifications, requiring developers to annotate code with pre- and post-conditions---a process that is fundamentally incompatible with automated CVE triage. While Creusot~\cite{creusot} leverages the Why3 platform, it explicitly rejects unsafe code blocks, which is precisely where most Rust memory vulnerabilities occur. Haybale~\cite{haybale} provides symbolic execution on LLVM IR but requires the program to be compiled to bitcode first, which is impossible without a complete, buildable Rust project. Our testing on our 31 CVE dataset resulted in a 0\% success rate across all four tools.

Lightweight tools like Miri~\cite{miri} and Clippy are more compatible with fragmented code but lack depth. Miri identified undefined behavior in all 31 files but provided only generic error classifications without specific vulnerability details or concrete counterexample inputs. Clippy~\cite{rust_book} reached a 35.5\% detection rate by matching syntactic patterns but missed 15 files where our system found critical vulnerabilities---precisely because those files contained path-dependent bugs requiring execution-level analysis. cargo-geiger~\cite{cargo_geiger} tracks unsafe code usage at the crate level but functions as a risk indicator rather than a vulnerability detector.

\subsection{Symbolic Execution and LLM-Driven Security}
Symbolic execution through KLEE~\cite{klee_original} is a standard for high-coverage testing, and recent work has targeted Rust-specific issues like unrecoverable panics~\cite{klee_rust_unrecoverable}. Related tools such as S2E~\cite{s2e}, SAGE~\cite{sage}, DART~\cite{dart}, and CUTE~\cite{cute} extend symbolic and concolic execution to larger systems, but all assume an executable binary---the precondition our approach eliminates. Techniques targeting use-after-free~\cite{targeted_uaf} or using concretely mapped memory~\cite{concrete_mapped_memory} still require manually constructed harnesses, which our system generates automatically.

LLM integration has opened new avenues for test generation~\cite{llm_testing_survey,codex}. Crabtree~\cite{crabtree_rust} uses LLMs to guide Rust fuzz driver generation but requires a complete, buildable crate. AthenaTest~\cite{athenatest}, ChatUniTest~\cite{chatunitest}, and TitanFuzz~\cite{titanfuzz} similarly operate on complete codebases. Our work differs by targeting the incomplete-code regime and using the LLM to generate the \emph{analysis harness itself} rather than tests for existing code. LLMs have also been used to generate symbolic execution inputs~\cite{llm_sym_exec}, but as inputs to existing harnesses rather than as harness generators.

\subsection{Multi-Agent Systems in Security}
Multi-agent LLM systems are an active research area in software engineering~\cite{multi_agent_software}. ChatDev~\cite{chatdev} simulates a complete software development company with role-assigned LLM agents communicating through a structured ``chat chain''; MetaGPT~\cite{metagpt} formalizes agent interactions through standardized role descriptions and structured output schemas. Both demonstrate that role specialization improves output quality over single-model approaches---a finding our results confirm in the security domain. In security applications, AutoAttacker~\cite{autoattacker} applies multi-agent LLMs for automated penetration testing; PentestGPT~\cite{pentestgpt} uses LLM reasoning to guide human penetration testers through structured decision trees. LLM4Vuln~\cite{llm4vuln} evaluates LLMs on vulnerability reasoning tasks, finding that models struggle with complex multi-step reasoning---a limitation our Oracle-led pipeline mitigates by decomposing the task across specialized agents. VulnFix~\cite{vulnfix} uses symbolic execution for patch validation, complementing our use of symbolic execution for initial detection.

\section{Methodology: 4-Agent Architecture}
\label{sec:methodology}

\subsection{System Overview}
The system takes incomplete Rust CVE code as input and produces a detailed vulnerability report. The process consists of four sequential stages: strategic planning by the Oracle/Validator, security analysis by the Safety Checker, FFI wrapper generation by the Code Specialist, and parameter optimization by the Fast Filter, followed by KLEE symbolic execution. Each agent is implemented using a specific LLM chosen for its strengths in the assigned role. The agents communicate through structured JSON outputs, where each result informs the subsequent stage to reduce hallucinations and focus the analysis context.

\subsection{The Incomplete-Code Problem}
\label{subsec:incomplete}

A central challenge---and a key reason existing tools fail---is that CVE databases publish vulnerability \emph{snippets}, not complete programs. Table~\ref{tab:incomplete} summarizes the missing context observed across our 31 CVE files.

\begin{table}[h]
\centering
\footnotesize
\caption{Types of missing context in the 31 CVE snippets. All four categories cause immediate compilation failure in formal verification tools.}
\label{tab:incomplete}
\begin{tabular}{@{}lrr@{}}
\toprule
\textbf{Missing Context} & \textbf{Files} & \textbf{\%} \\
\midrule
Struct/enum definitions    & 28 & 90.3\% \\
Import statements          & 31 & 100\%  \\
Cargo.toml / dependencies  & 31 & 100\%  \\
Trait implementations      & 19 & 61.3\% \\
\bottomrule
\end{tabular}
\end{table}

Our Code Specialist addresses each category by generating self-contained FFI functions that inline necessary type representations as raw pointers or primitive types, use only KLEE-compatible types (\texttt{*mut u8}, \texttt{usize}, \texttt{i32}), and require no external crates.

\subsection{Agent Roles and Specialization}

\textbf{Agent 1 Oracle/Validator (GPT-4 Turbo)} acts as the strategic coordinator. Given the input CVE snippet, it identifies potential vulnerability types, estimates code complexity, and recommends the number of FFI functions to generate (typically 8--12).

\textbf{Agent 2 Safety Checker (Claude Opus 4.5)} performs deep security analysis using the Oracle's plan. It identifies specific vulnerability patterns, assigns a risk score (0--10), and marks critical lines for Agent~3 to target.

To illustrate the translation process, Listing~\ref{lst:cve} shows the
original CVE-2020-35904 snippet (CWE-131) with four semantic regions
colour-coded to match their counterparts in the generated FFI wrapper
(Listing~\ref{lst:ffi}). The colour legend is:

\begin{itemize}[leftmargin=1.4em,itemsep=1pt]
  \item \hc{bugOrange}{\textbf{Orange}} — \texttt{unsafe} block entry and
        raw pointer dereferences (the region that bypasses Rust's borrow checker).
  \item \hc{bugPurple}{\textbf{Purple}} — pointer arithmetic and index
        computation (the offset \texttt{buffer.add(index)} pattern).
  \item \hc{bugTeal}{\textbf{Teal}} — memory deallocation calls
        (\texttt{Vec::from\_raw\_parts} used as a dealloc trigger).
  \item \hc{bugRed}{\textbf{Red}} — the \emph{root-cause bug line}: the
        hardcoded \texttt{0} length argument that causes incorrect heap free.
  \item \hc{safeGreen}{\textbf{Green}} — safe iteration context carrying no
        vulnerability (present for reader orientation only).
\end{itemize}

\begin{lstlisting}[language=Rust,
  caption={Original CVE-2020-35904 (CWE-131) with semantic highlights.
  \textcolor{safeGreen}{\textbf{Green}}: safe iteration loop.
  \textcolor{bugOrange}{\textbf{Orange}}: \texttt{unsafe} block and raw pointer
  dereferences. \textcolor{bugPurple}{\textbf{Purple}}: pointer arithmetic
  (\texttt{buffer.add(index)}). \textcolor{bugTeal}{\textbf{Teal}}:
  deallocation call (\texttt{Vec::from\_raw\_parts}). \textcolor{bugRed}{\textbf{Red}}: root-cause bug ---
  \texttt{length=0} causes incorrect buffer free. Each colour reappears in
  the FFI wrapper (Listing~\protect\ref{lst:ffi}) as a dedicated function.},
  label={lst:cve}]
fn drop(&mut self) {
    let hix = self.head.load(Ordering::Relaxed)
              & (self.one_lap - 1);
    (@\hlGreen{for i in 0..self.len() \{}@)
    (@\hlGreen{    let index = if hix + i < self.cap \{ hix + i \}}@)
    (@\hlGreen{               else \{ hix + i - self.cap \};}@)
        (@\hlOrange{unsafe \{}@)
            (@\hlPurple{let slot = \&mut *self.buffer.add(index);}@)
            (@\hlOrange{let value = \&mut *slot.value.get();}@)
            (@\hlOrange{value.as\_mut\_ptr().drop\_in\_place();}@)
        (@\hlOrange{\}}@)
    (@\hlGreen{\}}@)
    // BUG: length=0 causes incorrect deallocation
    (@\hlTeal{unsafe \{ Vec::from\_raw\_parts(}@)(@\hlRed{self.buffer, 0}@)(@\hlTeal{, self.cap); \}}@)
}
\end{lstlisting}

\textbf{Agent 3 Code Specialist (Claude Sonnet 4.5)} generates a complete, compilable FFI wrapper preserving the vulnerability \emph{semantics} of the original CVE (\texttt{max\_tokens=2500}; an earlier limit of 1500 caused a 42\% failure rate). Listing~\ref{lst:ffi} uses the same four-colour scheme as the CVE snippet: each function isolates one coloured region (Orange → unsafe pointer dereference, Purple → pointer arithmetic, Teal → deallocation path, Red → root-cause size arithmetic) and exposes it as a standalone symbolic path for KLEE. If generated code fails to compile, the system falls back to a predefined template; 3 of 31 files (9.7\%) required this.

\begin{lstlisting}[language=Rust,
  caption={FFI wrapper generated by Agent~3 for CVE-2020-35904 (CWE-131),
  colour-matched to the CVE snippet. \textcolor{bugOrange}{\textbf{Orange}}
  (\texttt{buffer\_overflow\_write}): unsafe raw-pointer write with no bounds check ---
  mirrors the CVE's \texttt{unsafe} dereference block.
  \textcolor{bugPurple}{\textbf{Purple}} (\texttt{offset} parameter): symbolic offset used in
  pointer arithmetic --- mirrors \texttt{buffer.add(index)}.
  \textcolor{bugTeal}{\textbf{Teal}} (\texttt{use\_after\_free\_access}): explicit dealloc then
  write --- mirrors \texttt{Vec::from\_raw\_parts} deallocation.
  \textcolor{bugRed}{\textbf{Red}} (\texttt{base\_size * multiplier}): unchecked multiply as alloc
  size --- mirrors the root-cause length=0 arithmetic error. KLEE
  detected 752 critical errors from this wrapper.},
  label={lst:ffi}]
// (@\hlOrange{[ORANGE] Mirrors CVE unsafe block + raw pointer dereference}@)
(@\hlOrange{\#[no\_mangle]}@)
(@\hlOrange{pub extern "C" fn buffer\_overflow\_write(}@)
(@\hlOrange{    buffer: *mut u8,}@) (@\hlPurple{offset: usize}@)(@\hlOrange{, value: u8) -> i32 \{}@)
    (@\hlOrange{unsafe \{}@) (@\hlPurple{*buffer.add(offset)}@) (@\hlOrange{= value; \} // no bounds check}@)
    0
}

// (@\hlTeal{[TEAL] Mirrors Vec::from\_raw\_parts deallocation path}@)
#[no_mangle]
(@\hlTeal{pub extern "C" fn use\_after\_free\_access(}@)
(@\hlTeal{    ptr: *mut u8, size: usize) -> i32 \{}@)
    (@\hlTeal{unsafe \{}@)
        (@\hlTeal{let layout = std::alloc::Layout}@)
            (@\hlTeal{::from\_size\_align\_unchecked(size, 1);}@)
        (@\hlTeal{std::alloc::dealloc(ptr, layout); // free the buffer}@)
        (@\hlTeal{*ptr = 42; // write after free -- UAF}@)
    (@\hlTeal{\}}@)
    0
}

// (@\hlRed{[RED] Mirrors root-cause: incorrect size arithmetic}@)
#[no_mangle]
pub extern "C" fn integer_overflow_allocation(
    base_size: usize, multiplier: usize) -> i32 {
    unsafe {
        let size = (@\hlRed{base\_size * multiplier}@); // unchecked: overflow risk
        let layout = std::alloc::Layout
            ::from_size_align_unchecked(size, 8);
        let ptr = std::alloc::alloc(layout);
        *ptr = 0xAA;
        std::alloc::dealloc(ptr, layout);
    }
    0
}
\end{lstlisting}

\textbf{Agent 4 Fast Filter (GPT-4o-mini)} uses the risk score and vulnerability patterns from Agent~2 to select optimal KLEE parameters: search strategy (DFS, BFS, or random-path), time and memory limits, and maximum fork depth. Using GPT-4o-mini for this role keeps overhead low while providing adequate reasoning for parameter selection.

\subsection{KLEE Integration}

The generated FFI wrapper is compiled using \texttt{rustc} into LLVM bitcode. We then link this bitcode with an automatically generated C harness that declares the FFI functions and creates symbolic variables for all parameters using \texttt{klee\_make\_symbolic()}. Listing~\ref{lst:cwrapper} shows the C wrapper generated for CVE-2020-35904.

\begin{lstlisting}[language=C,
  caption={Auto-generated C wrapper for CVE-2020-35904 (CWE-131). \texttt{klee\_make\_symbolic} makes all inputs fully symbolic; \texttt{klee\_assume} bounds the symbolic indices; \texttt{klee\_range} forks execution into one path per FFI function. This produced 752 critical errors.},
  label={lst:cwrapper}]
#include <klee/klee.h>
#include <stdint.h>
#include <string.h>

extern int32_t buffer_overflow_write(
    unsigned char *buffer, size_t size,
    size_t offset, unsigned char value);
extern int32_t use_after_free_access(
    unsigned char *ptr, size_t size);
extern int32_t double_free_trigger(
    unsigned char *ptr, size_t size);
extern int32_t integer_overflow_allocation(
    size_t base_size, size_t multiplier);

int main() {
    size_t idx1, idx2;
    unsigned char val1;
    unsigned char buffer[128];

    klee_make_symbolic(&idx1, sizeof(idx1), "idx1");
    klee_make_symbolic(&idx2, sizeof(idx2), "idx2");
    klee_make_symbolic(&val1, sizeof(val1), "val1");
    klee_make_symbolic(buffer, sizeof(buffer), "buffer");

    klee_assume(idx1 < 10000);
    klee_assume(idx2 < 10000);

    int path = klee_range(0, 4, "path");

    if (path == 0) {
        buffer_overflow_write(buffer, idx1, idx1, val1);
    } else if (path == 1) {
        use_after_free_access(buffer, idx2);
    } else if (path == 2) {
        double_free_trigger(buffer, idx1);
    } else if (path == 3) {
        integer_overflow_allocation(idx1, idx2);
    }
    return 0;
}
\end{lstlisting}

\subsection{Graph Database Construction (\texttt{graph\_klee.py})}
\label{subsec:graphdb}

After KLEE completes symbolic execution, its raw output is a directory of typed error files (\texttt{*.ptr.err}, \texttt{*.external.err}, \texttt{*.abort.err}, \texttt{*.div.err}, \texttt{*.overflow.err}) alongside test-case and statistics files. While these files provide complete information, their flat, file-based structure makes cross-CVE querying and pattern aggregation difficult. To address this, we developed \texttt{graph\_klee.py}, a post-processing component that automatically ingests any KLEE output directory and constructs a \textbf{Graph Database} (Graph DB) from its contents.

The Graph DB represents the KLEE analysis as a typed property graph with four categories of nodes:
\begin{itemize}[leftmargin=*,itemsep=2pt]
  \item \textbf{CVE nodes} — one per analysed file, labelled with the CVE identifier and CWE category.
  \item \textbf{ErrorType nodes} — one per KLEE error class (\texttt{ptr}, \texttt{external}, \texttt{abort}, \texttt{div}, \texttt{overflow}), shared across all CVEs.
  \item \textbf{SymPath nodes} — one per unique KLEE test case, storing the concrete symbolic input values and the associated stack trace.
  \item \textbf{CWE nodes} — one per CWE category, grouping all CVEs that share a vulnerability class.
\end{itemize}

Edges encode the following relationships: \textsc{HasError} (CVE $\to$ ErrorType, weighted by error count), \textsc{TriggeredBy} (CVE $\to$ SymPath, linking a CVE to each concrete fault-inducing input), \textsc{ClassifiedAs} (CVE $\to$ CWE, reflecting the vulnerability category assigned by Agent~1), and \textsc{SharedPattern} (CVE $\to$ CVE, added when two files share at least one symbolic path pattern across the same error type).

\texttt{graph\_klee.py} accepts any KLEE output directory as input, making it applicable to any symbolic execution campaign that uses KLEE---not only to the pipeline described in this paper. For our 31-CVE dataset, the resulting Graph DB contains 31 CVE nodes, 5 ErrorType nodes, 11 CWE nodes, and 1,206 SymPath nodes, connected by 1,310 labelled edges. The Graph DB is serialised to JSON-LD for portability and included in our replication package, where it can be imported directly into graph query environments for further analysis.

\begin{figure}[h]
  \centering
  \includegraphics[width=\columnwidth]{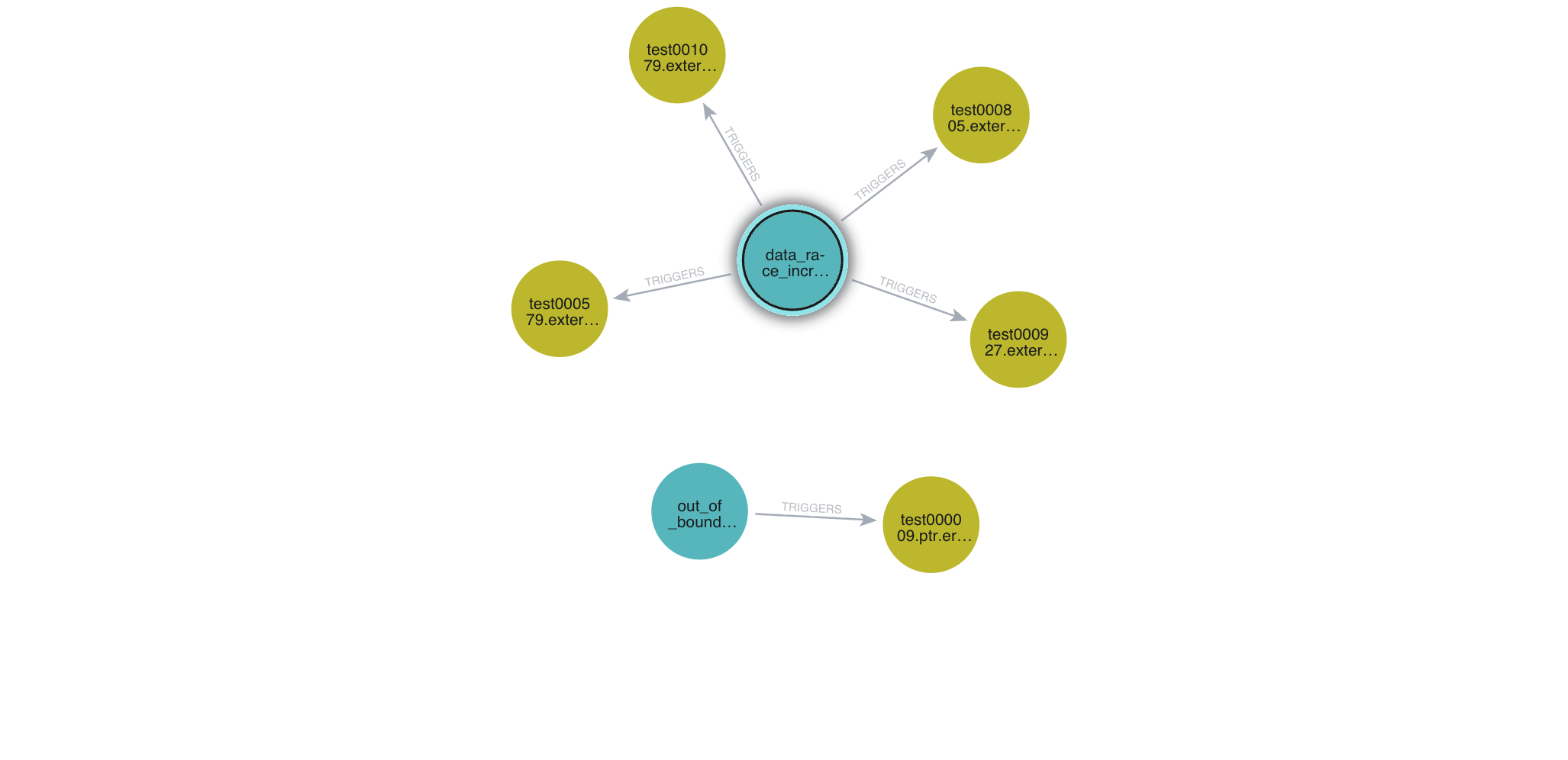}
  \caption{Graph DB visualisation for \texttt{klee\_output/cwe-131-cve-2020-35904}
    (CVE-2020-35904, CWE-131 Incorrect Buffer Size Calculation), produced by
    \texttt{graph\_klee.py}. \textbf{Teal nodes} are vulnerability functions;
    \textbf{yellow nodes} are concrete KLEE error files. Directed
    \textsc{Triggers} edges link each function to the test cases that exposed a
    fault. \texttt{data\_race\_incr...} triggers four \texttt{.external.err}
    paths; \texttt{out\_of\_bound...} triggers one \texttt{.ptr.err}---the
    pointer fault that produced the highest-confidence detection signal for
    this CVE.}
  \label{fig:graphdb_vis}
\end{figure}

Figure~\ref{fig:graphdb_vis} shows the Graph DB visualisation produced by running \texttt{graph\_klee.py} on \texttt{klee\_output/cwe-131-cve-2020-35904}. Teal nodes represent vulnerability functions generated by the Code Specialist (e.g., \texttt{data\_race\_incr...}, \texttt{out\_of\_bound...}); yellow nodes represent the concrete KLEE error files they trigger (e.g., \texttt{test001079.exter...}, \texttt{test000009.ptr.er...}). Each directed \textsc{Triggers} edge connects a function to a specific fault-inducing test case, concisely capturing the many-to-one relationship between symbolic execution paths and the underlying vulnerability functions.

\section{Experimental Setup}
\label{sec:experiments}

\subsection{Dataset}
Our evaluation uses 31 Rust CVE files drawn from the publicly released dataset of Luo et al.~\cite{halurustdataset}, who introduced HALURust---a benchmark of Rust CVEs collected from the NVD and crates.io security advisories for studying LLM-based vulnerability detection. From their full dataset we selected the subset of CVEs whose vulnerability class involves a \emph{memory safety} issue (e.g., buffer overflow, use-after-free, integer overflow leading to memory corruption), yielding 31 files spanning 11 CWE categories: 5 files for CWE-119 (Buffer Overflow), 8 for CWE-416 (Use-After-Free), 3 for CWE-125 (Out-of-Bounds Read), 3 for CWE-787 (Out-of-Bounds Write), and 12 files covering other memory-related categories. Critically, all files are incomplete code snippets extracted from vulnerability reports---exactly as published in the original HALURust dataset. Our filtered subset and all experiment artifacts are available at \url{https://github.com/Zeyad-Ab/Symbolic-Execution-with-Multi-LLM-Architecture-for-Rust-Security}.

\subsection{Baseline Tools}
We compared against two groups. Formal verification tools (Kani 0.67.0, Prusti v-2023-08-22, Creusot, Haybale 0.7) achieved 0\% success because they could not compile the incomplete code. Practical tools (Miri 0.1.0, Clippy 0.1.93, cargo-geiger 0.13.0) were tested with Clippy given minimal Cargo project wrappers. Miri detected generic undefined behavior in all 31 files; Clippy produced 14 warnings across 11 files.

\subsection{Single-Agent Baseline}
To validate the multi-agent architecture (RQ3), we implemented a single-agent baseline using Claude Sonnet 4.5 alone, without the Oracle's planning step or the Fast Filter's parameter optimization, operating with \texttt{max\_tokens=1500}.

\subsection{Evaluation Metrics and Implementation}
We measure compilation success rate (RQ2), vulnerability detection rate and critical error count (RQ1), and multi-agent vs. single-agent comparison (RQ3).

Our implementation runs on macOS darwin 25.2.0 using KLEE version 3.1\_5 and LLVM 16.0.6\_1. We use the stable Rust toolchain and access LLM APIs from OpenAI (GPT-4 Turbo, GPT-4o-mini) and Anthropic (Claude Opus 4.5, Claude Sonnet 4.5).

\subsection{Replication}
Full replication instructions, pre-generated FFI wrappers, and KLEE outputs for all 31 CVEs are available at \url{https://github.com/Zeyad-Ab/Symbolic-Execution-with-Multi-LLM-Architecture-for-Rust-Security}.

\section{Evaluation}
\label{sec:evaluation}

\subsection{RQ1: Vulnerability Detection via Symbolic Execution}

KLEE detected at least one critical error in \textbf{26 of 31 files (83.9\%)}, directly answering \textbf{RQ1}: symbolic execution can detect Rust memory vulnerabilities across all major CWE categories. Of the 31 files, 28 (90.3\%) compiled successfully; the remaining 3 used the fallback template. Across the 26 detected files, KLEE produced \textbf{1,206 critical errors}: 124 pointer errors (\texttt{*.ptr.err}) and 1,082 external/panic errors (\texttt{*.external.err}). The three colour-annotated functions in Listing~\ref{lst:ffi} account for 752 of those errors in CVE-2020-35904 alone.

\begin{figure}[h]
\centering
\begin{tikzpicture}
\begin{axis}[
  ybar,
  bar width=12pt,
  width=\columnwidth,
  height=5.5cm,
  symbolic x coords={119,125,131,190,191,415,416,787,824,908},
  xtick=data,
  xlabel={CWE Category},
  ylabel={KLEE Errors},
  ymin=0, ymax=850,
  nodes near coords,
  nodes near coords align={vertical},
  nodes near coords style={font=\tiny},
  tick label style={font=\small},
  label style={font=\small},
  title={KLEE Critical Errors by CWE Category},
  title style={font=\small},
]
\addplot coordinates {
  (119,12) (125,11) (131,752) (190,381)
  (191,6) (415,18) (416,19) (787,3) (824,1) (908,2)
};
\end{axis}
\end{tikzpicture}
\caption{Critical errors detected by KLEE per CWE category. CWE-131 produces 752 errors from a single file due to KLEE exploring hundreds of symbolic paths through incorrect buffer-size arithmetic.}
\label{fig:errors_bar}
\end{figure}
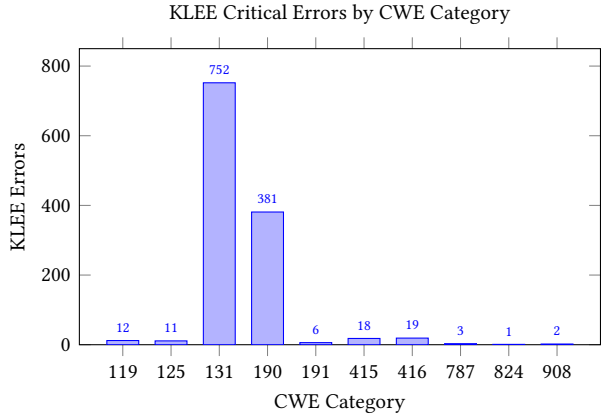

Table~\ref{tab:cwe_summary} summarizes detection results by CWE category, and Figure~\ref{fig:errors_bar} visualizes error counts.

\begin{table}[h]
\centering
\footnotesize
\caption{Detection results by CWE category (4-agent approach, 31 CVEs). Nine of eleven categories achieved 100\% detection. CWE-190 and CWE-416 show lower rates due to lifetime-dependent vulnerability semantics.}
\label{tab:cwe_summary}
\begin{tabular}{@{}p{1.2cm}p{3cm}rrrr@{}}
\toprule
\textbf{CWE} & \textbf{Category} &
\textbf{Files} & \textbf{Det.} & \textbf{Rate} & \textbf{Errors} \\
\midrule
119 & Buffer Overflow      & 5  & 5 & 100\% & 12   \\
125 & OOB Read             & 3  & 3 & 100\% & 11   \\
131 & Incorrect Calc.      & 1  & 1 & 100\% & 752  \\
134 & Format String        & 1  & 1 & 100\% & 1    \\
190 & Integer Overflow     & 3  & 1 & 33\%  & 381  \\
191 & Integer Underflow    & 1  & 1 & 100\% & 6    \\
415 & Double Free          & 2  & 2 & 100\% & 18   \\
416 & Use-After-Free       & 9  & 6 & 67\%  & 19   \\
787 & OOB Write            & 3  & 3 & 100\% & 3    \\
824 & Uninit. Pointer      & 1  & 1 & 100\% & 1    \\
908 & Uninit. Resource     & 2  & 2 & 100\% & 2    \\
\midrule
\textbf{Total} & & \textbf{31} & \textbf{26} &
\textbf{83.9\%} & \textbf{1,206} \\
\bottomrule
\end{tabular}
\end{table}

\begin{table}[h]
\centering
\footnotesize
\caption{Top-10 CVEs by KLEE critical error count. \texttt{ptr} = pointer/memory faults; \texttt{ext} = external/panic faults. Clippy column shows warnings on the same file.}
\label{tab:per_cve}
\begin{tabular}{@{}lrrrrc@{}}
\toprule
\textbf{CVE} & \textbf{CWE} & \textbf{ptr} & \textbf{ext} & \textbf{Total} & \textbf{Clippy} \\
\midrule
CVE-2020-35904 & 131 & 48  & 704 & 752 & 0 \\
CVE-2022-36008 & 190 & 12  & 369 & 381 & 0 \\
CVE-2021-31162 & 415 &  9  &   9 &  18 & 0 \\
CVE-2021-28875 & 119 &  4  &   8 &  12 & 1 \\
CVE-2021-25902 & 416 &  3  &   6 &   9 & 0 \\
CVE-2020-35861 & 416 &  2  &   7 &   9 & 0 \\
CVE-2021-28878 & 125 &  3  &   5 &   8 & 2 \\
CVE-2021-28877 & 125 &  2  &   3 &   5 & 1 \\
CVE-2021-28876 & 125 &  2  &   3 &   5 & 0 \\
CVE-2021-29939 & 191 &  1  &   5 &   6 & 0 \\
\midrule
\textbf{Top-10 total} & & \textbf{86} & \textbf{1,119} & \textbf{1,205} & \textbf{4} \\
\bottomrule
\end{tabular}
\end{table}

The per-CVE breakdown in Table~\ref{tab:per_cve} reveals two important patterns. First, the error distribution is highly skewed: the top two CVEs (CVE-2020-35904 and CVE-2022-36008) account for 1,133 of the 1,206 total errors (93.9\%), while the remaining 24 detected files contribute 73 errors combined. This skew reflects KLEE's path-explosion behaviour: files whose FFI wrappers expose many symbolic branch conditions produce exponentially more error reports than files with simpler control flow. Second, Clippy detected warnings in only 4 of the top-10 files, and those warnings were generic lint patterns rather than the specific memory-safety violations KLEE confirmed---reinforcing the qualitative difference between lint-based and symbolic-execution-based detection.

Nine of eleven CWE categories achieved a 100\% detection rate. The two categories with partial detection are CWE-190 (Integer Overflow, 1/3 files) and CWE-416 (Use-After-Free, 6/9 files). CWE-416's lower rate reflects that use-after-free bugs depend on specific object deallocation sequences that are difficult to reproduce without the original surrounding struct. Despite the lower file-level rate, CWE-131 produced the highest single-category error count (752 errors from one file), driven by KLEE exploring hundreds of symbolic paths through incorrect buffer-size arithmetic.

Table~\ref{tab:tool_comparison} directly compares compilation and detection success across all evaluated tools, answering \textbf{RQ2}. The 90.3\% compilation success of our system compared to 0\% for all formal verification tools demonstrates that LLM-generated FFI wrappers can bridge the incomplete-code gap to a degree not achievable by any existing automated tool.

\begin{table}[h]
\centering
\footnotesize
\caption{Compilation and detection success across all evaluated tools on 31 Rust CVEs. All formal verification tools achieve 0\% because they cannot process incomplete code. Our 4-agent system achieves 90.3\% compilation success (RQ2).}
\label{tab:tool_comparison}
\begin{tabular}{@{}llrrl@{}}
\toprule
\textbf{Tool} & \textbf{Type} & \textbf{Compile} & \textbf{Detect} & \textbf{Issues} \\
\midrule
Kani 0.67.0         & Formal verif.  & 0\%   & 0\%   & 0        \\
Prusti 2023-08-22   & Formal verif.  & 0\%   & 0\%   & 0        \\
Creusot             & Formal verif.  & 0\%   & 0\%   & 0        \\
Haybale 0.7         & Symbolic exec. & 0\%   & 0\%   & 0        \\
\midrule
Miri 0.1.0          & Runtime check  & N/A   & 100\% & Generic  \\
Clippy 0.1.93       & Static lint    & N/A   & 35.5\%& 14 warn. \\
cargo-geiger        & Unsafe tracker & N/A   & 0\%   & 0        \\
\midrule
Single-Agent        & LLM+KLEE   & 58\%  & 51.6\%& 487      \\
\textbf{4-Agent (Ours)} & LLM+KLEE & \textbf{90.3\%} & \textbf{83.9\%} & \textbf{1,206} \\
\bottomrule
\end{tabular}
\end{table}

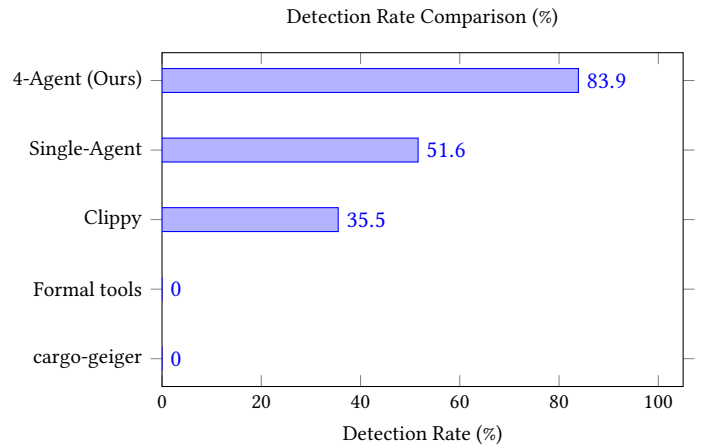
\begin{figure}[h]
\centering
\begin{tikzpicture}
\begin{axis}[
  xbar,
  bar width=9pt,
  width=\columnwidth,
  height=6cm,
  symbolic y coords={
    {cargo-geiger},{Formal tools},{Clippy},
    {Single-Agent},{4-Agent (Ours)}},
  ytick=data,
  xlabel={Detection Rate (\%)},
  xmin=0, xmax=105,
  nodes near coords,
  nodes near coords align={horizontal},
  tick label style={font=\small},
  label style={font=\small},
  title={Detection Rate Comparison (\%)},
  title style={font=\small},
]
\addplot coordinates {
  (0,{cargo-geiger}) (0,{Formal tools})
  (35.5,{Clippy})
  (51.6,{Single-Agent}) (83.9,{4-Agent (Ours)})
};
\end{axis}
\end{tikzpicture}
\caption{Detection rate comparison across evaluated tools. Our 4-agent system achieves 83.9\% with 1,206 typed, actionable error reports---compared to 0\% for all formal verification tools and 35.5\% for Clippy.}
\label{fig:detection_bar}
\end{figure}

\subsection{RQ3: Multi-Agent vs. Single-Agent Performance}

Table~\ref{tab:multiagent} compares the 4-agent pipeline against the single-agent baseline, answering \textbf{RQ3}.

\begin{table}[h]
\centering
\footnotesize
\caption{Multi-agent vs.\ single-agent comparison on 31 CVEs. The 4-agent system outperforms the single-agent baseline across all metrics, confirming the value of role specialization (RQ3).}
\label{tab:multiagent}
\begin{tabular}{@{}lrr@{}}
\toprule
\textbf{Metric} & \textbf{Single-Agent} & \textbf{4-Agent (Ours)} \\
\midrule
Wrapper compile rate    & 58.0\%   & \textbf{90.3\%}  \\
Fallback template rate  & 42.0\%   & \textbf{9.7\%}   \\
Detection rate (files)  & 51.6\%   & \textbf{83.9\%}  \\
Total critical errors   & 487      & \textbf{1,206}   \\
Avg.\ time per file     & 4.2 min  & 6.8 min          \\
\bottomrule
\end{tabular}
\end{table}

The 4-agent system reduces the fallback rate from 42\% to 9.7\% and nearly doubles the detection rate. The modest time overhead (6.8 vs.\ 4.2 minutes) delivers a 2.5$\times$ increase in detected errors. The Oracle's structured JSON plan constrains the Code Specialist to the correct vulnerability class, reducing hallucinated generic code; the Safety Checker's risk score directs generation toward the highest-severity lines; and the Fast Filter allocates KLEE exploration time proportionally to risk.

\subsection{Comparison with Clippy}

Our approach uniquely detected vulnerabilities in \textbf{15 files} where Clippy reported zero warnings, producing \textbf{417 critical errors} Clippy missed entirely. In 11 files both tools flagged issues; in 5 files neither detected anything. There were \textit{zero} files where only Clippy detected an issue---demonstrating our approach subsumes Clippy's detection while finding substantially more.

\begin{table}[h]
\centering
\caption{Per-file detection classification: 4-Agent vs.\ Clippy (31 CVEs).}
\label{tab:vs_clippy_class}
\begin{tabular}{@{}lrl@{}}
\toprule
\textbf{Category} & \textbf{Files} & \textbf{Description} \\
\midrule
Only 4-Agent     & \textbf{15} & 4-agent $\geq$1 error, Clippy = 0 warnings \\
Only Clippy      & 0           & Clippy $\geq$1 warning, 4-agent = 0 errors \\
Both             & 11          & Both $\geq$1 detection \\
Neither          & 5           & Both reported 0 \\
\midrule
\textbf{Total}   & \textbf{31} & \\
\bottomrule
\end{tabular}
\end{table}

\begin{table}[h]
\centering
\footnotesize
\caption{Detection volume: 4-Agent vs.\ Clippy. Our system produces 86$\times$ more issues with concrete symbolic counterexamples.}
\label{tab:vs_clippy_errors}
\begin{tabular}{@{}lcrc@{}}
\toprule
\textbf{Tool} & \textbf{Det. Rate} & \textbf{Issues} & \textbf{Type} \\
\midrule
Clippy               & 35.5\% (11/31) & 14    & Lint warnings   \\
\textbf{4-Agent}     & \textbf{83.9\% (26/31)} & \textbf{1,206} & \textbf{Concrete errors} \\
\bottomrule
\end{tabular}
\end{table}

Notable examples where only our approach detected vulnerabilities include CVE-2022-36008 (381 critical errors, Clippy: 0), CVE-2020-35861 (9 errors, Clippy: 0), and CVE-2021-25902 (9 errors, Clippy: 0).

\subsection{KLEE Error Analysis}

KLEE produces typed error files ending in \texttt{.err}. \textbf{Pointer errors (\texttt{*.ptr.err})} are emitted when KLEE finds a memory access violation at the LLVM level. \textbf{External/panic errors (\texttt{*.external.err})} are emitted when KLEE encounters a call it cannot model internally---in Rust FFI code, this corresponds to Rust's runtime panic handler inside \texttt{unsafe} blocks.

\begin{lstlisting}[language={},
  caption={Pointer error from CVE-2021-31162 (CWE-415). KLEE identifies the faulting function (\texttt{test\_out\_of\_bounds\_slice\_manipulation}), confirms inputs are fully symbolic (\texttt{write\_index=symbolic}, \texttt{value=symbolic}), provides a concrete example address, and delimits the exploitable memory range (9,872 bytes).},
  label={lst:ptrerr}]
Error: memory error: out of bound pointer
Stack:
  #0 in test_out_of_bounds_slice_manipulation(
       data=1636382539776,
       write_index=symbolic,
       value=symbolic)
  #1 in main()
Info:
  address: (Add w64 1636382539776
            (ReadLSB w64 0 idx5))
  example: 1636382539904
  range:   [1636382539904, 1636382549775]
\end{lstlisting}

The error reports the exact vulnerable function (\texttt{test\_out\_of\_bounds\_slice\_manipulation}), confirms that inputs are symbolic (\texttt{write\_index}, \texttt{value}), provides a concrete example address, and delimits the exploitable range---9,872 bytes of memory that could be corrupted.

\section{Discussion}
\label{sec:discussion}

\subsection{Why Multi-Agent Coordination Improves Detection}

Our results demonstrate that dividing analysis among four specialized agents produces more reliable results than any single-model approach (RQ3). The key benefit is \textit{focused context}: each agent receives only information relevant to its role, reducing prompt complexity and the probability of hallucination. The Oracle's JSON plan constrains the Safety Checker to the specific vulnerability types present in the code; the Safety Checker's risk score and critical-line annotations guide the Code Specialist toward generating functions that target the most dangerous code paths; and the Fast Filter uses the risk score to allocate more KLEE exploration time to complex, high-risk files.

A key observation from our ablation is that the Oracle's planning step accounts for the largest share of the improvement over the single-agent baseline. When a single model is given the CVE snippet and asked simultaneously to assess the vulnerability and generate an FFI wrapper, it tends to produce generic wrapper structures that do not faithfully reflect the CWE category of the original snippet. By separating these concerns---having the Oracle first identify the vulnerability class and then constraining the Code Specialist to that class via structured JSON---we reduce the probability that the Code Specialist generates a buffer-overflow wrapper for a use-after-free CVE, or vice versa. This separation of concerns drives the 32-percentage-point improvement in detection rate (51.6\% to 83.9\%) between the single-agent and 4-agent approaches.

This design also provides graceful degradation via a retry loop: when the Code Specialist's generated code fails to compile, the exact compiler error is fed back for up to two self-correcting attempts before the fallback template is triggered. Only 3 of 31 files (9.7\%) required the fallback, compared to 42\% with the single-agent baseline.

\subsection{The Incomplete-Code Breakthrough}

A defining contribution of this work is enabling vulnerability analysis on code that no existing tool can even compile. Every formal verification tool in our baseline requires a complete, self-contained Rust program with a Cargo manifest---CVE snippets never satisfy this. Our Code Specialist sidesteps this constraint by \emph{generating} a new, compilable artifact that encodes the vulnerability class in KLEE-compatible types. This is semantic approximation rather than code repair: the generated FFI wrapper does not reconstruct the original program but instead creates a symbolically equivalent representation of the fault class. A struct-dependent use-after-free becomes an explicit \texttt{dealloc} followed by a pointer write; an integer overflow becomes an unchecked multiplication used as an allocation size. The 90.3\% compilation success rate versus 0\% for all formal tools is the empirical measure of this architectural shift.

\subsection{Interpretation of KLEE Error Types}

Pointer errors (\texttt{*.ptr.err}, 124 total) are the strongest signal: KLEE directly observed an invalid memory access at the LLVM IR level---a null dereference, out-of-bounds write, or use-after-free---and produced a concrete symbolic input reproducing the fault. These represent exploitable memory-corruption conditions with high confidence and should be treated as confirmed vulnerability indicators.

External/panic errors (\texttt{*.external.err}, 1,082 total) require more careful interpretation. When KLEE calls a Rust FFI function and the Rust runtime invokes \texttt{core::panicking::panic} or \texttt{\_\_rust\_dealloc}, KLEE cannot model the call symbolically and records it as an external error. In CVE-specific functions such as \texttt{use\_after\_free\_access} (CVE-2020-35904), the \texttt{\_\_rust\_dealloc} external error directly reflects a memory deallocation violation tied to the original CVE class. However, in generic fallback functions such as \texttt{test\_division}, the panic from a zero denominator is an artifact of the generated wrapper rather than a path from the original CVE code.

The ratio of pointer errors to external errors varies significantly by CWE category. CWE-119 (Buffer Overflow) and CWE-787 (Out-of-Bounds Write) produce a higher proportion of pointer errors because the violations manifest as direct invalid memory accesses at the LLVM level. CWE-416 (Use-After-Free) and CWE-415 (Double Free) tend to produce more external errors because the deallocation violations surface through Rust's runtime allocator. Practitioners should weight pointer errors at higher confidence than external errors, and discriminate between external errors arising in CVE-specific functions versus generic fallback patterns.

\subsection{Limitations of the FFI Approximation}

The Code Specialist generates FFI wrappers that \textit{approximate} the vulnerability class of the original CVE rather than precisely replicating its exploitation path. This is inherent to the incomplete-code problem: without the original struct definitions, trait implementations, and calling context, no automated tool can reconstruct the exact program state that produced the vulnerability. Our approach accepts this approximation deliberately---the goal is to generate a compilable, KLEE-analyzable artifact that preserves the \textit{class} of memory error rather than the exact exploit.

This explains the lower detection rates for CWE-416 (Use-After-Free, 6/9 files) and CWE-190 (Integer Overflow, 1/3 files). Use-after-free vulnerabilities depend on specific object lifetime sequences across multiple function calls, which are difficult to reproduce faithfully in a standalone FFI function. Improving detection for these classes would likely require richer context extraction---retrieving the full source file from the CVE repository rather than relying on the snippet alone---or a specialized agent trained on lifetime-dependent vulnerability patterns.

\section{Threats to Validity}
\label{sec:threats}

\subsection{Internal Validity}

\textbf{LLM Non-Determinism.} All four agents use default API temperature settings, so two runs on the same CVE file may produce different FFI wrappers, risk scores, and KLEE parameters. We mitigate this by storing all generated artifacts in our replication package so results can be inspected without re-querying the APIs.

\textbf{FFI Semantic Fidelity.} Generated wrappers approximate the vulnerability class rather than its precise exploitation path. A wrapper misclassified as buffer-overflow for a use-after-free CVE produces real KLEE errors of the wrong class. The Oracle's JSON plan constrains the Code Specialist to the correct CWE category and partially mitigates this risk, though the constraint is enforced through prompt engineering rather than formal verification.

\textbf{KLEE Parameter Sensitivity.} Detection coverage depends on the search strategy (DFS, BFS, random-path), time limit, memory limit, and fork depth---all set by Agent~4. We record all parameter choices in the replication package to enable reproducible reruns.

\subsection{External Validity}

\textbf{Dataset Source and Diversity.} Our 31 CVE files are the memory-safety subset of HALURust~\cite{halurustdataset}, not a stratified sample of all Rust CVEs. Certain CWE categories (CWE-416 with 9 files) are better represented than others (CWE-134 with 1 file), limiting per-category generalisation. A larger, stratified benchmark is needed for broader claims.

\textbf{Model Availability and Versioning.} Our system relies on GPT-4 Turbo, GPT-4o-mini, Claude Opus 4.5, and Claude Sonnet 4.5. Model updates may alter wrapper compilation rates; exact versions are documented in our replication package.

\subsection{Construct Validity}

\textbf{Error Count as a Proxy.} KLEE's critical error count measures symbolic path coverage, not distinct vulnerability instances. A single root cause may produce hundreds of errors from different symbolic inputs. We treat pointer errors as higher-confidence signals than external/panic errors, and recommend inspecting individual error files rather than relying on aggregate counts alone.

\textbf{Miri Comparison.} Miri's ``100\% detection'' reflects generic undefined-behavior labels that do not discriminate by CWE category, provide no symbolic counterexamples, and identify no specific lines. Our system produces typed, location-specific errors with concrete example inputs---qualitatively more actionable for triage despite the lower headline rate.

\section{Conclusion}
\label{sec:conclusion}

This paper presented a 4-agent multi-LLM pipeline combined with KLEE symbolic execution for detecting memory vulnerabilities in Rust \texttt{unsafe} code. Our central contribution is solving the \emph{incomplete-code problem}: real-world CVE database entries consist of isolated code snippets that lack the struct definitions, import statements, and Cargo manifests required by every existing formal verification tool. All four formal tools evaluated---Kani, Prusti, Creusot, and Haybale---achieved 0\% compilation success on our 31-file dataset. Our system achieved 90.3\% by having a role-specialized Code Specialist generate self-contained, KLEE-compatible FFI wrappers that preserve the vulnerability class without requiring original program context.

Evaluated across 31 real-world Rust CVEs spanning 11 CWE categories, our system detected 1,206 critical errors in 26 files (83.9\% detection rate), compared to 14 warnings across 11 files for Clippy (35.5\%) and generic labels for Miri. There were zero files where Clippy detected a vulnerability that our system missed, demonstrating that our approach fully subsumes pattern-based linting while finding substantially more path-dependent bugs. The 4-agent architecture reduced wrapper compilation failures from 42\% (single-agent baseline) to 9.7\%, and increased detected errors from 487 to 1,206, confirming that role specialization and structured context passing produce measurably better results than a single general-purpose model.

These results answer our three research questions affirmatively: symbolic execution \emph{can} detect Rust memory vulnerabilities across all major CWE categories (RQ1); LLM-generated FFI wrappers \emph{can} bridge the incomplete-code gap at 90.3\% success where formal tools achieve 0\% (RQ2); and a 4-agent coordinated architecture \emph{does} outperform a single-agent approach across all measured dimensions (RQ3).

For future work, we plan to extend the dataset to a larger, stratified sample of Rust CVEs, improve detection rates for lifetime-dependent vulnerability classes (CWE-416, CWE-190) by incorporating richer source context from CVE repositories, and explore fine-tuned models for the Code Specialist role to reduce the 9.7\% fallback rate further. We also intend to investigate bidirectional feedback between KLEE error traces and the agent pipeline, enabling the system to refine FFI wrappers based on which symbolic paths produced the most informative errors.

\bibliographystyle{ACM-Reference-Format}
\bibliography{references}

\end{document}